\documentstyle[12pt]{article}
\title{\ \\ Shock waves in the dissipative Toda lattice}
\author{Jarmo Hietarinta
\\Department of Physics, University of Turku\\FIN-20500 Turku, Finland\\
\\
Tom Kuusela
\\Department of Applied Physics, University of Turku\\
FIN-20500 Turku, Finland\\ and\\
Boris A. Malomed\\Department of Applied Mathematics,
School of Mathematical Sciences\\ Raymond and Beverly Sackler Faculty of Exact
Sciences\\Tel Aviv University, Ramat Aviv 69978, Israel}

\begin{document}
\begin{titlepage}
\renewcommand{\thepage}{}

\maketitle

\begin{abstract}
We consider the propagation of a shock wave (SW) in the damped Toda
lattice. The SW is a moving boundary between two semi-infinite lattice
domains with different densities. A steadily moving SW may exist if
the damping in the lattice is represented by an ``inner'' friction,
which is a discrete analog of the second viscosity in hydrodynamics.
The problem can be considered analytically in the continuum
approximation, and the analysis produces an explicit relation between
the SW's velocity and the densities of the two phases. Numerical
simulations of the lattice equations of motion demonstrate that a
stable SW establishes if the initial velocity is directed towards the
less dense phase; in the opposite case, the wave gradually spreads
out.  The numerically found equilibrium velocity of the SW turns out
to be in a very good agreement with the analytical formula even in a
strongly discrete case. If the initial velocity is essentially
different from the one determined by the densities (but has the
correct sign), the velocity does not significantly alter, but instead
the SW adjusts itself to the given velocity by sending another SW
in the opposite direction.
\end{abstract}

\end{titlepage}

\baselineskip=20pt

\section{Introduction}
The analysis of dynamical behavior in linear and nonlinear lattices is
of obvious interest for various branches of solid state physics
\cite{Mar,Toda}. It is known that anharmonic interactions in the
lattice may produce stable localized collective excitations in the
form of solitons \cite{Toda,others} in the absence of
dissipation. The dissipation naturally damps solitons \cite{damping},
although they may be supported, in certain cases, by an external AC
drive \cite{ac}.

The aim of the present work is to consider a different nonlinear
wave in the damped lattice, namely a shock wave (SW) or kink that is
realized as a localized transient region between two parts of the
lattice with different spacings, i.e., different densities. Evidently,
the shock may be produced by an initial deformation of the lattice.
Another way to produce a shock is to apply a strong short pulse to an
edge of the lattice (e.g., by an impact of another body).
Mathematically, the latter type of the shock corresponds to the von
Neumann problem: in an undeformed lattice, the particles with the
positive and negative numbers are given initial velocities $\pm V_0$
\cite{von}. A similar but different problem was considered by Kaup
\cite{Kaup} for the semi-infinite lattice, in which the first
particle is driven at a constant velocity, while others are initially
at rest.

For the non-dissipative (exactly integrable) Toda lattice (TL) model,
evolution of the von Neumann shock has been analyzed in a rather full
detail \cite{exact}. It has been demonstrated analytically and
numerically that the evolution is essentially non-stationary, producing
a lot of oscillations behind the shock.  (This seems to be
qualitatively similar to the decay of an initial step configuration in
the Korteweg -- de Vries (KdV) equation, which is a continuum limit of
the TL \cite{Pit}.) The principal difference between the von Neumann's
problem and the problem to be considered in this work is that the
former one is initially dealing with a homogeneous lattice and
infinitely large {\em kinetic} energy (if the system is infinite),
while we want to consider a lattice that is initially deformed so that
it has an infinite reservoir of {\em potential} energy.

The presence of suitable dissipation in the lattice seems to change
the behavior in an essential way: With dissipation it may be possible
to obtain a steadily propagating shock between the phases (lattice
domains) with different densities.  It is interesting that the
inclusion of friction, which renders the problem physically relevant,
also is necessary for the {\em stable} propagation of certain waves.

In this work we will consider shock propagation in the framework of
the damped TL although the general results obtained below should
remain qualitatively correct for a fairly broad class of nonlinear
dynamical models. In the TL one can introduce two different types of
friction, which can be naturally called outer and inner. The model
with the outer friction is based on the following lattice equation of
motion:
\begin{equation}
\ddot{x} _n+\gamma \dot{x}_n \, =\, e^{-\left( x_{n}-x_{n-1}\right) }
-e^{-\left(x_{n+1} -x_n\right) }\, ,
\end{equation}
where $x_n$ is a displacement of the $n$th particle from its
equilibrium position. The dissipative term in Eq.\ (1) implies
friction between the lattice and some external substrate, hence the
name ``outer'' friction.  However, for applications to solid state
physics, it seems more natural to consider friction of purely internal
origin, which is similar to the second viscosity in hydrodynamics,
i.e., resistance against change of the density of the lattice. The TL
model with inner friction is based on the following equation of
motion:
\begin{equation}
\ddot{x}_n+\gamma \left( 2\dot{x}_n-\dot{x}_{n-1}-\dot{x}_{n+1}\right)\,
=\, e^{-\left( x_n-x_{n-1}\right) } -e^{-\left( x_{n+1} -x_n\right) }\, .
\label{E:xi}
\end{equation}
A local density of the lattice is determined by the local relative
displacement $r_n\equiv x_n-x_{n-1}$, which in the inner friction case
obeys the equation
\begin{equation}
\ddot{r}_n+\gamma \left( 2\dot{r}_n-\dot{r}_{n-1}-\dot{r}_{n+1}\right)\,
=\, 2e^{-r_n} -e^{-r_{n+1}}-e^{ -r_{n-1} }.
\label{E:ri}
\end{equation}
The variables $x_n$ in Eqs. (1) and (\ref{E:xi}) are
not necessarily full coordinates of the particles, but may be, as it
was mentioned above, displacements of the particles from some
regularly spaced (with distance $a$) equilibrium positions. The full
distance between the particles will then be $a+r_n$, and the local
density of the lattice $(a+r_n)^{-1}$. This also implies that, in
what follows below, $r_n$ may take both positive and negative values.

In this paper our main concern is the study of SW's governed by Eq.
(\ref{E:ri}).  In the second section we obtain some analytical results
using the continuum approximation. We can demonstrate, e.g, that a
steadily traveling SW always exists, and although it cannot be
generally found in an explicit form, we obtain an analytical
expression for its velocity, which turns out to depend on the
asymptotic values of $r$ in front of and behind the SW, but not on the
friction coefficient $\gamma$ in Eq.\ (\ref{E:ri}).  This velocity is
sandwiched between the sound velocities of the two phases, and the
motion is towards the less dense phase. In the third section, we
display results of numerical simulations of Eq.\ (\ref{E:ri}).  When
the initial velocity of the SW has the correct sign, we always end up
with a stably propagating wave of a permanent shape, but when the
direction is wrong the SW gradually flattens out.  We find also that
when the SW is started with a wrong magnitude of the velocity (but
with the correct sign), the SW readjusts its {\em height} so that a
fairly good agreement is reached with the analytically predicted
formula. The change of the height is carried out by emitting an
additional SW in the opposite direction

\section{Analytical consideration}
As the first step in analyzing the possibility of a steady propagation
of the SW, it is natural to define the ``quasi-momentum''
\begin{equation}
P \, \equiv \, \sum_{n=-\infty}^{n=+\infty} \dot{x}_n\, .
\end{equation}
Summing up all the equations (1) over $n$ from $-\infty$ to
$+\infty$, one immediately obtains the equation
\begin{equation}
\frac{dP}{dt} +\gamma P\, =\, 0\, .
\label{E:quasi}
\end{equation}
Since the friction coefficient $\gamma$ is positive, Eq.\
(\ref{E:quasi}) tells us that $P\to 0$ as $t\to \infty$. On the other
hand, it is obvious that a steadily propagating SW must have a
nonzero value of the quasi-momentum.  Therefore, a steadily
propagating SW is not possible in the model (1).

Proceeding to the model (\ref{E:xi}) with the inner friction, one notices,
first of all, that the same trick yields, instead of Eq.\
(\ref{E:quasi}), the equation $\frac{d P} {dt} =0$. Thus, in this case
a constant nonzero value of the quasi-momentum is possible.

Let us now consider the continuum approximation.  Assume, as usual,
that $r_n$ in (\ref{E:ri}) changes appreciably only on a scale much
larger than the lattice spacing. One may treat $n$ as a
quasi-continuous variable (coordinate), and then in the lowest
approximation, Eq.\ (\ref{E:ri}) goes over into the continuum equation
\begin{equation}
r_{tt} -\gamma\, r_{tnn} \, =\, -\left( e^{-r} \right)_{nn} \, ,
\label{E:rcon}
\end{equation}
where the subscripts stand for the corresponding partial derivatives.
Note that, generally speaking, we do not assume the value of $r_n$
small, and therefore we do not expand the exponential.
For the time being we do not take into account the fourth derivative
on the right-hand side. In what follows below it will be added to
Eq.\ (\ref{E:rcon}) when necessary.

One should now look for traveling-wave solutions to
Eq.\ (\ref{E:rcon}) in the form $r=r(z), \, z\equiv n-Vt$, $V$ being
the wave's velocity. Substituting this into Eq.\ (\ref{E:rcon}), one
arrives at an ordinary differential equation which can be integrated
twice. With regard to the condition that we are interested in
solutions which remain finite at $n\rightarrow \pm \infty$, the final
form of the integrated equation is
\begin{equation}
\gamma V\frac{dr}{dz} \, =\, -\left( V^2r+e^{-r}-C\right) \,
\label{E:coneq}
\end{equation}
$C$ being an arbitrary constant of integration.  In what follows, we
will consider the asymptotic values $r_{\pm}\equiv r(z=\pm \infty ) $ as
given parameters (as it was explained above, they are determined by
the density of the lattice in front of and behind the SW), and
since the derivative $\frac{dr}{dz}$ must vanish at the two
infinities, we find
\begin{eqnarray}
V^2&=&\left( r_{+}-r_{-}\right)^{-1}\left(e^{-r_-}-e^{-r_+}\right),
\label{E:vel}\\
C&=&\left(r_{+}-r_{-}\right)^{-1}\left(r_+ e^{-r_- }-r_-e^{-r_+}\right).
\label{E:velC}
\end{eqnarray}
Note that the friction coefficient does not enter in these relations.
For the maximum value of the slope we get
\begin{equation}
\max\left(\frac{dr}{dz}\right)=\frac1{\gamma V}\left(V^2\log V^2-V^2+C\right),
\label{E:maxslope}
\end{equation}
this value is reached when $r=-\log V^2$.

The linearization of Eq.\ (\ref{E:rcon}), with $\gamma =0$,
around an arbitrary constant value $r_0$ ($r=r_0+\rho$, $\rho$ being a
small variable part of $r$) yields a D'Alembert equation for $\rho$,
which determines the sound velocity $s$ in the homogeneous lattice
with $r=r_0$:
\begin{equation}
s^2\, =\, e^{-r_0} \, .
\label{E:svel}
\end{equation}
It is easy to check that the shock's velocity given by Eq.\
(\ref{E:vel}) always lies between the sound velocities corresponding
to $r=r_{\pm}$, and in the limit $r_+-r_-\rightarrow 0$ coincides with
the corresponding sound velocity. Furthermore, the velocity
(\ref{E:vel}) coincides with the sound velocity at the maximum slope
(\ref{E:maxslope}).

An explicit profile of the SW cannot be obtained from Eq.\
(\ref{E:coneq}) in a closed form. However, this can be done in the
limiting case $|r_+-r_-|\ll r_{\pm}$. In this case, Eq.\
(\ref{E:coneq}) reduces to
\begin{equation}
2\gamma s^{-1}\frac{dr}{dz} \, =\, (r_+-r)(r-r_-) \, ,
\label{E:limeq}
\end{equation}
where $s$ is the sound velocity (\ref{E:svel}) corresponding to
$r_{\pm}$ (in this approximation, we neglect the difference between
them). The solution to Eq.\ (\ref{E:limeq}) is a typical kink
\begin{equation}
r(z)\, =\, \frac{1}{2}\left[ (r_+-r_-)+(r_+-r_-)\,\tanh \zeta \right] \, ,
\end{equation}
where $\zeta \equiv \frac{1}{4} s\gamma ^{-1} (r_+-r_-)z$.

Let us now briefly consider the applicability conditions for the
approximation considered. As it follows from the explicit solution
(\ref{E:limeq}), the size $L$ of the SW can be estimated as
follows:
\begin{equation}
L \, \sim \, \gamma /V\, \delta r\,
\end{equation}
where $\delta r \equiv r_+-r_-$. Thus, the basic condition for
application of the continuum approximation, $L \gg 1$, takes the form
\begin{equation}
\gamma \, \gg \, V\delta r\, .
\label{E:concon1}
\end{equation}

Above, we have neglected the fourth derivative on the right-hand side
part of Eq.\ (\ref{E:rcon}). This is justified if, expanding the
exponential, one has the second derivative of the squared variable part
$\rho$ of $r$ much larger than the fourth derivative $\rho
_{nnnn}$. Using again the solution (\ref{E:limeq}), one concludes that
the corresponding condition is
\begin{equation}
\gamma ^2 \, \gg \, V^2\delta r\, .
\label{E:concon2}
\end{equation}
In the general case, when $\delta r$ is not specially small, the
conditions (\ref{E:concon1}) and (\ref{E:concon2}) are actually
equivalent, i.e., there is no necessity to add the fourth derivative
to Eq.\ (\ref{E:rcon}). However, in the case of the {\it weak} SW,
when $\delta r$ is small, the inequality (\ref{E:concon1}) is less
restrictive than (\ref{E:concon2}).

Representing, as above, $r$ in the form $r_0+\rho$, one obtains an
equation for the (small) variable part $\rho$.  This equation can be
immediately integrated once, so that one ends up with the known
Korteweg-de Vries - Burgers equation \cite{Karp}:
\begin{equation}
2\rho _t-\gamma \rho _{zz}-s\rho \rho _z+\frac{1}{12} s\rho _{zzz} \,
= \, 0\, ,
\label{E:KdVB}
\end{equation}
where $z$ is the same traveling coordinate as in Eq.\ (\ref{E:coneq}),
and $s$ is the sound velocity (\ref{E:svel}) corresponding to $r=r_0$.
In the work \cite{Karp} it has been demonstrated that Eq.\
(\ref{E:KdVB}) has steady traveling-wave solutions in the form of a
SW with an oscillating tail (it was approximately described in
[9] as a bound state of several KdV solitons). This solution
represents the lattice SW in this limiting case.

Also in this case an exact analytical solution for the shock's shape
is not available, but one can easily find its velocity from
Eq.\ (\ref{E:KdVB}).  Going back to the original coordinate $n$, one
can eventually obtain the full shock's velocity as follows:
$V=s-\frac{1}{4} s(\rho _++\rho_-)$, where $\rho _{\pm} \equiv r_{\pm}
-r_0$. On the other hand, expansion of the general formula
(\ref{E:vel}) for small $r_+-r_-$ yields exactly the same
result. Thus, the addition of the fourth derivative affects the shape
of the SW, but not its velocity.

To conclude this section, it is relevant to note that the continuum
approximation has another limit of applicability when the lattice
density in front of the SW becomes too small, i.e., $r_+$ is very
large.  Formally, Eq.\ (\ref{E:vel}) predicts that the velocity
vanishes in this limit. However, it follows from Eq.\ (\ref{E:coneq})
that the size of the corresponding SW is, in the continuum
approximation, $L \sim \gamma Vr_-e^{r_-}$. Thus, $L$ shrinks when $V$
vanishes, and the continuum approximation is no longer valid.

\section{Numerical results}
We have simulated Eq.\ (\ref{E:ri}) numerically in the logarithmic form
\begin{equation}
\partial^2_t \ln(1+v_n) + \gamma\partial_t \left[2\ln(1+v_n)-\ln(1+v_{n+1})
-\ln(1+v_{n-1})\right]=v_{n+1}+v_{n-1}-2 v_n ,
\label{E:vform}
\end{equation}
where $v_n=\exp(-r_n)-1$, or more precisely, in its two-component form
\begin{eqnarray}
\partial_t\, v_n &=& (i_n-i_{n+1})(1+v_n), \label{E:viformv} \\
\partial_t\, i_n &=& v_{n-1}-v_n - \gamma (2i_n-i_{n+1}-i_{n-1}).
\label{E:viformj}
\end{eqnarray}
System (\ref{E:viformv},\ref{E:viformj}) is more suitable for
numerical integration than the original one, because there is no need
to evaluate transcendental functions, and therefore local numerical
errors can be significantly reduced. We have used the Bulirsch-Stoer
algorithm as the integration method: the total number of the lattice
sites was 2000. With a lattice of this length it was possible to study
long-term evolution of the initial state.

As the initial state we took $v_n(0)$ where $v_n(t)$ is a SW of the form
\begin{equation}
v_n(t)= v_0+\frac{1}{2}\Delta v \left( 1 - \tanh[k(n-n_0-Vt)]\right),
\label{E:viiniv}
\end{equation}
and then $i_n(0)$ was obtained from (\ref{E:viformj}) with
(\ref{E:viiniv}) and $\gamma=0$ by integrating in time:
\begin{equation}
i_n(0)= {\displaystyle \frac{\Delta v}{2Vk}\ln\left(
\frac{\cosh[k(n-1-n_0)]}{\cosh[k(n-n_0)]}\right)}.
\label{E:viinij}
\end{equation}
[Note that $i_n$ is only needed up to a constant additive part.]  In
these formulae the velocity $V$ and slope $k$ are free, but it turns
out, that if we want the best initial value we should take $V$ from
(\ref{E:vel}) with $r_+=-\ln(1+v_0+\Delta v)$ and $r_-=-\ln(1+v_0)$,
and the slope of the initial state from (\ref{E:maxslope}) (recall
$r_n=-\log(v_n+1)$):
\begin{equation}
k = -\frac{2\; V}{\Delta v\gamma}
\left(V^2\ln V^2 - V^2 +C\right).
\end{equation}

The time evolution of a typical solution is shown in Figs.\ 1(a) and
1(b) in the form of snapshots of the lattice at various times. If the
dissipation is zero (Fig.\ 1(a)) the front of the SW starts
immediately to produce oscillatory ripple waves at its top shoulder.
These oscillations increase in amplitude and extend slowly to the left
of the shoulder. If we have, however, moderate dissipation ($\gamma=1$
in Fig.\ 1(b)), the SW travel with constant shape and velocity
without any oscillating tail. We have varied the amplitude parameters
$v_0$ and $\Delta v$ from $0.5$ to $10$, and the dissipation factor
$\gamma$ from $0.1$ to $2$ (which implies that the width of the slope
varies from 2 to 50 lattice points) and always obtained similar
results.  Typical final SW profiles are presented in Fig.\ 2
using three different values for the dissipation factor. The velocity
of the SW can be predicted extremely well (deviation is less than
0.2\%) using the continuum limit result (\ref{E:vel}), even when the
system is highly discrete.

The convergence of the initial solution towards the final steady state
is illustrated in Fig.\ 3(a). This figure contains several snapshots
of the lattice plotted on top of each other, after a certain shift
relative to the lattice. To determine the shift, we first sought for
the position of the SW in the last snapshot by numerically fitting to
the (theoretical) solution (\ref{E:viiniv},\ref{E:viinij}) with
constant initial amplitude parameters $v_0$ and $\Delta v$. The waves
in the other snapshots were then shifted using only the theoretical
velocity and the evolved time.  The slope of the initial state was on
purpose made much smaller than the correct one. The first three
snapshots have been plotted with open circles, triangles and squares,
and the remaining ones with solid circles. One can see that the
initial state changes its slope and converges rapidly to the final
common shape. It should be noted that this process does not generate
any noticeable radiation, this is because the dissipation is strongest
in the region which dominates the generation of radiation.

The initial form (\ref{E:viiniv},\ref{E:viinij}) is actually very
close to the final traveling wave solution. This is demonstrated in
Fig.\ 3(b), where we have shown the difference between the theoretical
approximation and the numerical result presented in Fig.\ 3(a). The
deviation is only few percent of the original shape.

If we change the sign of $i_n$ in (\ref{E:viinij}), the SW starts to
travel in the opposite direction as shown in Fig.\ 4. In this case,
the profile of the wave is not constant, but rather its slope
decreases gradually.

The most interesting feature of the SW is how it adjusts itself when
the initial velocity is different from that given by Eq.\
(\ref{E:vel}), this is illustrated in Fig.\ 5. If $i_n$ is smaller
than the correct value (i.e., the velocity is too large, c.f.\ Eq.
(\ref{E:viinij})), the system adjusts locally the upper level $r_+$
to agree with the given velocity (Fig.\ 5(a)), and this, in turn,
gives rise to another SW at the top level. This new SW travels away
from the original SW, but since this is the wrong direction this SW
slowly flattens out.  If, instead, we use initially $i_n(0)$ that is
larger than the right one, the upper level of the wave increases
(Fig.\ 5(b)), and the two SWs propagate in the opposite
directions. For both the shocks, the velocities have the correct sign,
therefore they quickly reach a stationary form without changing the
slope (Fig.\ 5(b)). It is remarkable that all these major changes take
place without any visible radiation.

\section{Conclusions}

In this paper we have studied the existence of stationary SW's in the
Toda lattice. Stable traveling waves of this type were found to exist
when the friction in the lattice is of the ``inner'' type as per Eq.\
(\ref{E:ri}). This is not accidental. From Fig.\ 1 we see that,
without dissipation, the system develops violent oscillations at the
upper ``shoulder'' (level) of the SW. The second spatial derivative of
the waveform has its maximum there as well, so that its time
derivative provides a sufficiently strong balancing dissipative force
according to Eq. (3).

With the inner dissipation, the SW quickly attains its final
stationary form. Using the continuum approximation, we have obtained
the formula (\ref{E:vel}) which relates the velocity of the SW to
the asymptotic values of the lattice density in front of and beyond
the SW. This formula proves to be very accurate even in the
strongly discrete case.

A very interesting feature of the SW is seen when the initial
velocity and the asymptotic values do not obey Eq. (\ref{E:vel}).
The SW will then adjust itself to the velocity, and this always seems to
happen through changing the upper level of the SW. This process gives
rise to a new SW, which is traveling in the opposite direction.

The existence of the stable SWs in the damped Toda lattice is a novel
feature of the system and it is very likely that it may be observed in
experiments.

\section*{Acknowledgments}
This work was initiated when one of the authors, B.A.M., was visiting
the University of Turku, supported by a joint grant from the Center
for International Mobility (Finland) and the Israeli Ministry of
Science and Technology.

\newpage

\section*{Figure captions}

\begin{itemize}

\item[Fig. 1:]
(a) The time evolution of the initial state given by Eq.\
(\ref{E:viiniv},\ref{E:viinij}) in the absence of dissipation
$(\gamma=0)$. The time interval is $100$, $v_0=1.0$, $\Delta v=1.0$,
$k=0.159$. For better visibility we have left out the 9th and 10th
snapshots. In this and all other figures we plot $v_n\equiv
\exp(-r_n)-1$; (b) the same as Fig. 1(a) but with dissipation
$\gamma=1.0$.

\item[Fig. 2:]
The final waveforms of three numerical solutions with different values
of the dissipation factor $\gamma$ (but with the same asymptotic densities
obtained from $v_0=1.0$, $\Delta v=0.5$).

\item[Fig. 3:]
(a) the approach to the stationary shape of the SW when the initial
state has an incorrect slope (the time interval is $50$, $v_0=1.0$,
$\Delta v=3.0$, $k=0.200$, $\gamma=0.5$).  Each curve has been shifted
relative to the lattice in order to center them at the same lattice
position (see the text). The three first curves are depicted, by open
circles, triangles and squares, respectively, the remaining ones by
solid circles; (b) the same as Fig. 3(a) but with the analytical
approximation (\ref{E:viiniv}) subtracted from each curve.

\item[Fig. 4:]
Snapshots of the lattice when the sign of initial $i_n(0)$ (hence the
sign of $V$) is ``incorrect'', the initial state being in the rear
($v_0=1.0$, $\Delta v=3.0$, $\gamma=0.5$).

\item[Fig. 5:]
(a) The time evolution of an initial state with an ``incorrect'' value
of the initial velocity (but with the correct sign) ($v_0=1.0$,
$\Delta v=3$, $\gamma=0.5$). In this case, the ``correct'' velocity
$V$ is $1.809$, while we set initially $V=10.0$; (b): the same but
with the initial velocity $V=1.0$.

\end{itemize}

\end{document}